\documentstyle[aps,graphicx,multicol,psfrag,citesort]{revtex}

\begin{document}

\bibliographystyle{prsty}

\title {\large \bf Minimal Model for Sand Dunes}

\author{Klaus Kroy\thanks{present address:
 Physics Dept., University of Edinburgh, King's
Buildings, Edinburgh EH9 3JZ, Scotland}, Gerd Sauermann, and Hans
J. Herrmann}

\address{PMMH, \'Ecole Sup.\ de Physique et Chimie
  Industrielles. 10, rue Vauquelin, 75231 Paris, Cedex 05,
  France.}

\maketitle

\begin{abstract}
  We propose a minimal model for aeolian sand dunes. It combines an
  analytical description of the turbulent wind velocity field above
  the dune with a continuum saltation model that allows for saturation
  transients in the sand flux. The model provides a qualitative
  understanding of important features of real dunes, such as their
  longitudinal shape and aspect ratio, the formation of a slip face,
  the breaking of scale invariance, and the existence of a minimum
  dune size.
\end{abstract}

\pacs{PACS numbers: }

\begin{multicols}{2} 
  
  Sand dunes develop wherever sand is exposed to an agitating medium
  (air, water \dots) that lifts grains from the ground and entrains
  them into a surface flow. The diverse conditions of wind and of sand
  supply in different regions on Earth give rise to a large variety of
  shapes of aeolian dunes
  \cite{Bagnold41,Pye90,Lancaster95}. Moreover, dunes have been found
  on the sea--bottom and even on Mars.  Despite the long history of
  the subject, the underlying physical mechanisms of dune formation
  are still not very well understood.  How are aerodynamics
  (hydrodynamics) and the particular properties of granular matter
  acting together to create dunes?  How is the shape of a dune
  maintained when it moves? In the following we propose a ``minimal
  model'' for aeolian sand dunes to address such questions.  Although
  it refers only to rather generic properties of the wind velocity
  field and the laws of aeolian sand transport, it can make
  interesting qualitative predictions that are not sensitive to the
  simplifying assumptions, e.g.\ about the surface profile, the
  development and position of the slip face, dune migration etc.
  Using results from turbulent boundary layer calculations
  \cite{jackson-hunt:75,hunt-leibovich-richards:88,weng-etal:91}, we
  will propose an approximate analytical description of the surface
  shear stress exerted by the wind onto a heap of sand.  This will be
  combined with a saltation model \cite{sauermann-kroy-herrmann:tbp}
  that allows for saturation transients of the sand flux, which are an
  essential element of a consistent description of dunes.
  
  We start by summarizing some basic knowledge about aeolian sand
  transport and saltation. The mean turbulent wind velocity above a
  plane surface increases logarithmically with height. Its magnitude
  is specified by a characteristic velocity called the ``shear
  velocity'' $u_*$ and defined by $u_*^2\equiv\tau_0/\rho$ with
  $\tau_0$ the average surface shear stress (far away from any
  obstacle) and $\rho$ the density of air.  On a surface covered with
  sand, the wind entrains some grains into a surface layer flow if the
  shear velocity exceeds a threshold value.  The grains advance mainly
  by an irregular hopping process, thereby reducing the wind velocity
  in the surface layer. A unique relation between the shear stress
  $\tau$ and the sand flux $q$ is thus established in the equilibrium
  state.  If $\tau$ is not too close to the threshold, one has
  approximately \cite{Bagnold41,Anderson88,Pye90,ActaMechanica91}
  \begin{equation} 
  \label{eq:satflux}
      q_s \propto \tau^{3/2}\;.
  \end{equation} 
  The index $s$ emphasizes that this simple relation is
  restricted to situations where the flux is saturated.
  According to Eq.(\ref{eq:satflux}), the changing wind shear stress
  above a heap of sand $h(x,y)$ is responsible for flux gradients,
  which cause erosion and deposition. Due to mass conservation, the
  flux gradient $dq/dx$ on a slice $h(x)$ parallel to the wind
  direction gives rise to a change in height with time $dh/dt$, and
  thus to migration of the surface profile in the wind direction with a
  velocity 
  \begin{equation} \label{eq:v_d} 
   v\propto dq/dh \;.
  \end{equation} 
  This velocity is much smaller than the velocity of
  the saltating grains, so that the terrain can be assumed to be
  stationary for considerations concerning the wind and saltation
  dynamics.  To obtain $q_s$ and the corresponding $v_s\equiv v(q_s)$
  from Eqs.(\ref{eq:satflux}) and (\ref{eq:v_d}), we thus need to know
  the stationary surface shear stress perturbation $\hat \tau \equiv
  \tau/\tau_0-1$ caused by the heap. This depends on its shape.
  
  In a first attempt to model $\hat \tau$, one might try the affine
  relation $\hat\tau \propto h$ and combine it with the saturated flux
  approximation $q=q_s$.  If this was true, the surface velocity $v$
  would increase with height ($dv/dh \geq 0$) due to the nonlinearity
  of Eq.(\ref{eq:satflux}).  This in turn would decrease the upwind
  slope and increase the lee slope, and thus eventually create a slip
  face on the lee side, where sand slides down in avalanches.  Since
  there is so far nothing in the model to stop the further decrease of
  the upwind slope, our model ``dune'' would then start to decrease in
  height and finally become flat, whatever its windward profile.  This
  ``zeroth order model'' is obviously missing an essential
  ingredient. In the following, we will successively demonstrate how
  the modeling of both the shear stress and the flux have to be
  refined to achieve a consistent description of dune formation and
  migration. In particular, we will show that a subtle balance of two
  quantitatively small but qualitatively crucial corrections to the
  zeroth order model determine the shape of aeolian heaps and
 \begin{figure}[t] 
  \psfrag{x}{$x$}
  \psfrag{h}[lr]{$f$} 
  \psfrag{t}[lr]{$\tau/\tau_0$} 
  \begin{center}
  \leavevmode \narrowtext
  \includegraphics[width=0.8\columnwidth]{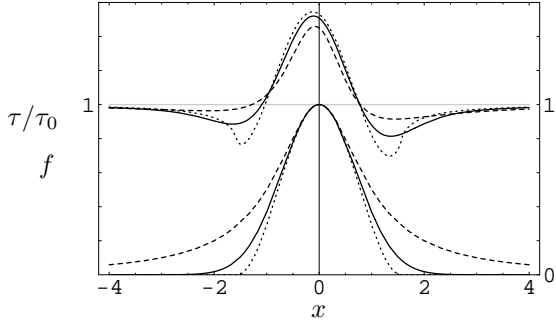} 
  \caption{Surface shear stress $\tau(x)/\tau_0$ (upper curves)
	above the profiles of Eq.(\ref{eq:hills}) (lower curves: $f_L$
	dashed, $f_G$ solid, $f^2_C$ dotted line) according to
	Eq.(\ref{eq:wind_model}) with $\alpha H/\!  L=0.45$,
	$\beta=0.25$. The small windward shift of the maximum shear
	stress with respect to the apex is crucial for dune
	formation.} \label{fig:hills} 
 \end{center} 
 \end{figure} \noindent
 dunes.

  An improved description for the shear stress can be
  obtained from turbulent boundary layer calculations. The surface
  shear stress perturbation $\hat \tau_x$ in wind direction caused by
  a smooth bump $h(x,y)$ is
  \cite{hunt-leibovich-richards:88,weng-etal:91} 
  \begin{equation}
  \label{eq:hunt} {\cal F}_{xy}\{\hat\tau_x\} = \frac{\alpha k_x( k_x
  + i \beta |k_x|)}{(k_x^2 + k_y^2)^{1/2}} \, {\cal F}_{xy} \{ h(x,y)
  \} \;.  
  \end{equation} 
  Here, we have neglected a minor term $\propto \ln |k_xL|$, and
  abbreviated the Fourier transformation from $x$, $y$ to $k_x$, $k_y$
  by ${\cal F}_{xy}$.  The parameters $\alpha$ and $\beta$ depend
  logarithmically on the ratio $L/z_0$, where $L$ is the
  characteristic length of $h(x,0)$ and $z_0$ is a measure of the
  surface roughness \cite{logs}. For the following discussion we evaluate
  Eq.(\ref{eq:hunt}) on a slice $h(x)\equiv h(x,0)$ of a transverse dune
  ($h$ independent of $y$). We
  obtain ($h':=dh/dx$) 
  \begin{equation}
  \label{eq:wind_model} \hat\tau(x) = \alpha \left[
  \int_{-\infty}^{\infty}\!\!\!\!  d\xi \; \frac{h'(x-\xi)}{\pi \xi} +
  \beta h'(x)\right] \;.  
  \end{equation} 
  Minor corrections for dunes with a finite transverse width
  can be calculated \cite{finite_width} but are omitted here for simplicity.
 
  For certain heap profiles $ h(x)\equiv H f(x/\! L)$
  (Fig.~\ref{fig:hills}) with
\begin{equation}
  \label{eq:hills}
  f_L = \left(1+x^2\right)^{-1},\;\; 
  f_G = e^{-x^2}\,,\;\; 
  f^n_C = \cos^{n} x  \,, 
\end{equation}
Eq.(\ref{eq:wind_model}) can be evaluated analytically. (The cosine
profile $f^n_C$ is understood to vanish outside the region
$[-\pi/2,\pi/2]$.)  The plots of Eq.(\ref{eq:wind_model}) in
Fig.~\ref{fig:hills} share several crucial properties not present in
the affine approximation $\hat\tau\propto h$.  First, we observe from
a dimensional analysis of Eq.(\ref{eq:wind_model}) that
$\hat\tau\propto H/\! L$ for a given shape $f$.  Hence, as one expects
from the scale invariance of turbulence, the amplification of the
shear velocity at the top of the profile is determined by its aspect
ratio $H/\!L$ and is essentially independent of the absolute height
$H$ \cite{logs}.  Secondly, at the tails of the profiles in
Fig.~\ref{fig:hills}, the shear stress decreases below its asymptotic
value $\tau_0$.  This effect is particularly pronounced for the
profile $f^n_C$ that has a discontinuous second derivative.  Further,
as a consequence of the second term in Eq.(\ref{eq:wind_model}), the
surface shear stress is asymmetric
\begin{figure}[t]
  \psfrag{v}[lr]{$v_s\; [a.u.]$}
  \psfrag{x}{$x$}
  \begin{center} \leavevmode  \narrowtext
    \includegraphics[width=0.9\columnwidth]{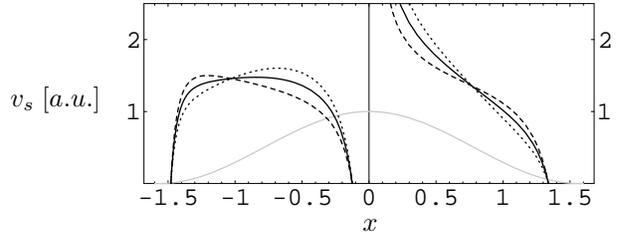} 
    \caption{The surface
      velocity $v_s(x)$ of the cosine--shaped heap $f^2_C(x)$ (gray)
      according to Eqs.(\ref{eq:satflux}), (\ref{eq:v_d}),
      (\ref{eq:wind_model}) for $\alpha=3.2$, $\beta=0.25$, and varying aspect
      ratios $H/\! L= 0.01$ (dashed), $0.14$ (solid), $0.27$ (dotted).  Note
      the tendency to increase the (unstable) lee slope and to drive the
      windward slope towards a (stable) optimum value that depends on $\alpha$
      and $\beta$. If $f^2_C$ is interpreted as the envelope of a dune and its
      separation bubble, the slip face must be located near the sharp drop of
      $v_s(x)$ slightly upwind from the top of the envelope.}
    \label{fig:v_d}
  \end{center}
\end{figure} \noindent
 even for symmetric profiles like those of Eq.(\ref{eq:hills}). For
$q=q_s$, there can thus be a net deposition on a symmetric heap of
sand and the shift of the shear stress maximum with respect to the top
of the heap \cite{lorentz} allows deposition at the top of the heap.
An initially flat heap of sand can spontaneously grow and
increase its windward slope.

To understand this instability of a plane sand surface, the migration
velocity $v_s(x)$ of a cosine shaped heap of sand $f^2_C(x)$ is
depicted in Fig.~\ref{fig:v_d}.  The decrease of $v_s(x)$ on the lee
side reveals the anticipated self--amplifying instability of the lee
slope, which gives rise to the formation of the slip face.  More
interestingly, Eq.(\ref{eq:wind_model}) renders $v_s(x)$ approximately
constant over almost the whole windward side if $H/\!L$ is close to a
stable optimum value.  Slightly better results can be achieved for
slightly lower $n$ (with slightly larger $H/\!L$) but not for the
profiles $f_G$ and $f_L$, for which $v_s$ is always non--uniform. The
dashed and dotted lines in Fig.~\ref{fig:v_d} were obtained for a
smaller and a larger aspect ratio and represent a steepening
($v_s'<0$) and flattening ($v_s'>0$) of the windward side,
respectively. Altogether, Fig.~\ref{fig:v_d} suggests that the coupled
Eqs.(\ref{eq:satflux}) and (\ref{eq:wind_model}) drive a heap of sand
towards a ``dune'' with a cosine--like windward profile of a preferred
aspect ratio, and a slip face on the lee side.

\begin{figure}[t]
  \begin{center} 
    \leavevmode \narrowtext
    \includegraphics[width=0.9\columnwidth,height=0.3\columnwidth]{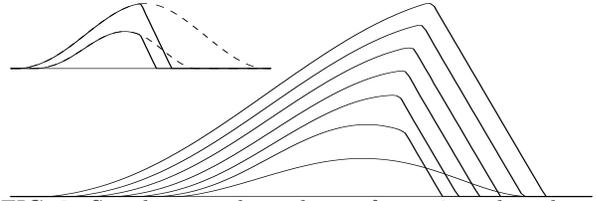}
\caption{Steady state dune shapes for various dune lengths (roughly
$50\,\ell_s \dots 100\,\ell_s$) obtained from Gaussian heaps by
integrating Eqs.(\ref{eq:v_d}) and (\ref{eq:flux}). Flow separation at
the brink was modeled heuristically by introducing a separation bubble
(dashed lines in the inset).  Periodic boundary conditions were
applied to conserve the sand mass and thus allow for a steady
state. The influx/outflux vanishes for the dunes with slip face but is
finite for the heap. The time evolution of the latter is shown in
Fig~\ref{fig:growth}b.  (Plot rescaled for visualization.)}
\label{fig:shapes}
\end{center}
\end{figure}

From the preceding qualitative analysis it is not yet obvious that
this process converges to a translation invariant steady state
solution.  Several previous studies using similar descriptions either
did not scrutinize the long time behavior of their models
\cite{Wippermann86,van_dijk-arens-boxel:99}, or failed to obtain
stable dunes \cite{zeman-jensen:88,Stam97}. Indeed, the present model
is still insufficient.  The problem with the quasi--linear
Eqs.(\ref{eq:hunt}) and (\ref{eq:wind_model}) is that they were
derived for smooth hills and cannot account for flow separation. Once
a slip face has developed, with a slope of about
$32^{\circ}-35^{\circ}$ that terminates in a sharp brink, they can no
longer be applied.  As in the textbook example of a backward facing
step, flow separation occurs at the brink and large eddies develop in
the wake region. The recirculating flow in this ``separation bubble''
is bounded by a separating streamline against the region of laminar
flow. Following a suggestion by Zeman and Jensen
\cite{zeman-jensen:88}, we assume that in order to calculate the shear
stress on the windward side, Eqs.(\ref{eq:hunt}) and
(\ref{eq:wind_model}) may be applied to the envelope of a dune and its
separation bubble (inset of Fig.~\ref{fig:shapes}) in place of the
actual surface profile of the dune.  In our numerical computations we
model this effective separation bubble by a third order polynomial
that is the smooth continuation of the windward dune profile with a
maximum slope of $14^{\circ}$ \cite{bubble}.

If this generalization is accepted, and if --- for qualitative
purposes --- one identifies the profile $f_C^2$ with such an envelope,
one can draw further interesting conclusions from Fig.~\ref{fig:v_d}.
First, the constant $v_s(x)$ obtained for the windward side of the
cosine profiles $f^n_C$ ($n \approx 2$), provides strong evidence that
Eqs.(\ref{eq:satflux}) and (\ref{eq:wind_model}) allow a dune with a
cosine--like windward side of suitable aspect ratio to remain
invariant upon translation by aeolian sand transport.  The steep
decrease of $v_s(x)$ at a certain distance upwind from the top of the
envelope, which is a consequence of the maximum of $\tau(x)$ being
shifted with respect to the top of the envelope, must then coincide
with the position of the brink.  Secondly, the stable dune shape is
scale invariant \cite{logs}.  Together with the scaling relation
$\hat\tau \propto H/\!L$ mentioned above, this implies that the
migration velocity $v$ of dunes obeys $v\propto u_*^3\!/\!  L $, in
accord with the empirical observation of a roughly reciprocal relation
between dune velocity and dune height \cite{Bagnold41,Pye90}.

However, one has to bear in mind that Eq.(\ref{eq:satflux}) is
restricted to a completely saturated saltation layer. Saturation
transients in the sand flux, as they occur under variable wind or sand
conditions, introduce a new characteristic length, the saturation
length $\ell_s$ \cite{sauermann-kroy-herrmann:tbp} related to (but
distinct from) the mean saltation length of the grains.  Shape
invariance is therefore restricted to large dunes
($L/\ell_s\to\infty$).  Estimates \cite{sauermann-kroy-herrmann:tbp}
of typical absolute values of $\ell_s$ suggest that real dunes are not
necessarily large in this sense, in accord with field measurements
\cite{sauermann-etal:2000} that report size dependent dune shapes.
Eq.(\ref{eq:satflux}) also fails at boundaries sand/ground.  It
predicts deposition at the windward foot of an isolated dune, where
the shear stress decreases \cite{Wiggs96,sauermann-kroy-herrmann:tbp}
(Fig.~\ref{fig:v_d}).  Previous numerical studies tried to avoid this
by focusing onto the short time behavior and by \emph{ad hoc}
heuristic methods such as a ``smoothing operator'' \cite{Wippermann86}
or an ``adaptation length'' \cite{van_dijk-arens-boxel:99}.

In the following, we demonstrate how these problems are naturally
cured if Eq.(\ref{eq:satflux}) is replaced by a slightly more
elaborate description derived from a continuum saltation model
\cite{sauermann-kroy-herrmann:tbp}.  In the steady state
($\partial/\partial t\simeq 0$), that model reduces to the
differential equation
\begin{equation}
  \label{eq:flux}
  \ell_s\partial q/\partial x = q( 1-q/q_s)  
\end{equation}
for the sand flux $q$, and explicit expressions for $\ell_s(\tau)$ and
$q_s(\tau)$. For wind velocities well above the entrainment threshold, 
$\ell_s(\tau)$ levels off, and the saturated flux $q_s$
\noindent 
converges to Eq.(\ref{eq:satflux}). The detailed forms used in our
numerical integrations, which involve the entrainment threshold, can
be found in \cite{sauermann-kroy-herrmann:tbp} but are not important
here. During the integration of Eqs.(\ref{eq:v_d}) and
(\ref{eq:flux}), the lee slope of the heap is constantly evaluated.
If it exceeds $14^{\circ}$, the separation bubble is introduced for
the shear stress calculation as described above.  Additionally, the
surface shear stress is set to zero within the separation zone, and
surface avalanches are induced if the slope exceeds the angle of
repose ($34^{\circ}$).  For definiteness we choose the values
$\alpha=3.2$, $\beta = 0.25$ for the parameters of
Eq.(\ref{eq:wind_model}).  Starting from a Gaussian heap, we obtain
the steady state profiles shown in Fig.~\ref{fig:shapes}. Their
general shape and average windward slope of about $10^{\circ}$ are in
good accord with recent field measurements
\cite{Hesp98,sauermann-etal:2000}. The shape of the largest dune
(indeed well described by $f_C^n$, $n\approx 2$) nicely confirms our
expectations from Fig.~\ref{fig:v_d}, whereas for small heaps, the
instability of the lee slope is washed out by the saturation
transients. The upwind shift of $\tau(x)$ with respect to $h(x)$
(which is of the order of $\beta L$) and the lag of $q(x)$ with
respect to $\tau(x)$ and $q_s(x)$ (of the order of $\ell_s$) compete
to establish a critical heap size for \emph{slip face formation}.
Smaller heaps do not spontaneously develop a slip face, while somewhat
larger heaps evolve into a steady state dune with a windward slope
that is lowered and a slip face that is shifted downwind compared to
the asymptotic shape.  The critical dune size depends on the wind
velocity through (and scales as) the saturation length $\ell_s(\tau)
\approx \ell_s(\tau_0)$ ($\approx$ for large $\tau$).

\begin{figure}[t]
  \begin{center} 
    \leavevmode \narrowtext
    \includegraphics[width=0.9\columnwidth]{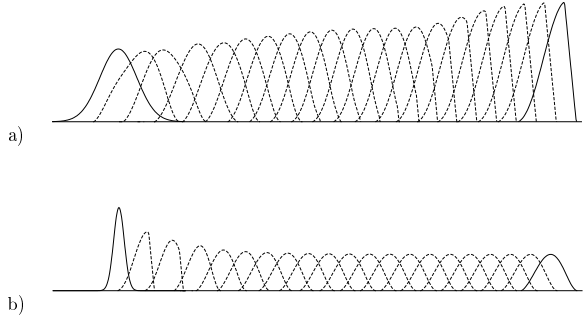}
    \caption{Growth histories for two Gaussian heaps of different mass
      and initial aspect ratio (periodic boundary conditions).
      The steep initial configuration of the heap in part b) gives
      rise to the temporary formation of a slip face, which turns out
      to be unstable.  Note that the distances to reach the steady
      state shape are different.  (Plots rescaled for visualization.)}
      \label{fig:growth} \end{center}
\end{figure}

Fig.~\ref{fig:growth} shows the growth histories of two heaps of
different mass and initial aspect ratio $H/\!L$. It demonstrates the
spontaneous formation of a slip face that is stable for the larger
heap but finally turns out to be unstable for the smaller one. For
intermediate sizes, flow separation can stabilize an existing slip
face although it would not have developed from a flat configuration.
This is the case for the two smallest dunes with slip face in
Fig.~\ref{fig:shapes}, which were produced from a steep initial
configuration (as in Fig.~\ref{fig:growth}b). As well as the critical
heap size for slip face 
\noindent
formation (out of a flat initial condition), there exists thus a
somewhat smaller critical dune size for \emph{slip face destruction}
and an intermediate regime, where the presence of a slip face depends
on the initial conditions. This implies, in particular, that a slip
face that develops out of a flat initial condition has always a finite
height.  Indeed, we find that slip face destruction also occurs at a
finite height.

In summary, we proposed a minimal model for aeolian sand dunes. The
advantage of the simple Eqs.(\ref{eq:hunt}) and (\ref{eq:wind_model})
over more elaborate nonlinear models
\cite{sykes:80,zeman-jensen:87,boxel-arens-van_dijk:99} for the wind
shear stress is that they combine a high computational efficiency with
good accuracy \cite{weng-etal:91}, and can thus be applied to simulate
large scale desert topographies.  The drawback is that we had to
introduce the heuristic separation bubble \cite{bubble}.  We pointed
out that the saturation transients contained in Eq.(\ref{eq:flux})
break the expected scale invariance and are responsible for the
existence of a critical heap size for slip face formation and a
critical dune size for slip face destruction. Further, we analyzed the
mechanisms responsible for the longitudinal shape and aspect ratio of
dunes and heaps, and for slip face formation.  It is an interesting
question whether (and how) these results can be generalized to full
$3d$ problems.  The deviations of the wind velocity from the main wind
direction are small above an obstacle that is flat in the transverse
direction.  One can thus expect that the present approach applies to
transverse dunes and, with minor corrections to
Eq.(\ref{eq:wind_model}) \cite{finite_width}, in the vicinity of the
symmetry plane of a crescent--shaped barchan dune.  As a direct
consequence of our above results, only $3d$ heaps larger than a
certain minimum size can spontaneously develop a slip face, which then
has a finite height and width.  This minimum size should scale as
$\ell_s$, which is a decreasing function of $u_*$
\cite{sauermann-kroy-herrmann:tbp}.  Preliminary results from an
extension of the model to $3d$ support this picture.

We gratefully acknowledge financial support by the DFG (HE
2732/1-1), and helpful discussions with J. Soares Andrade Jr.\ and
M. E. Cates.


\begin{thebibliography}{10}

\bibitem{Bagnold41}
R.~A. Bagnold, {\em The physics of blown sand and desert dunes.} (Methuen,
  London, 1941).

\bibitem{Pye90}
K. Pye and H. Tsoar, {\em Aeolian sand and sand dunes} (Unwin Hyman, London,
  1990).

\bibitem{Lancaster95}
N. Lancaster, {\em Geomorphology of desert dunes} (Routledge, London, 1995).

\bibitem{jackson-hunt:75}
P.~S. Jackson and J.~C.~R. Hunt, Q. J. R. Meteorol. Soc. {\bf 101},  929
  (1975).

\bibitem{hunt-leibovich-richards:88}
J.~C.~R. Hunt, S. Leibovich, and K.~J. Richards, Q. J. R. Meteorol. Soc. {\bf
  114},  1435  (1988).

\bibitem{weng-etal:91}
W.~S. Weng {\it et~al.}, Acta Mech. Suppl. {\bf 2},  1  (1991).

\bibitem{sauermann-kroy-herrmann:tbp}
G. Sauermann, K. Kroy, and H.~J. Herrmann, Phys. Rev. E {\bf 64}, 031305  (2001).

\bibitem{Anderson88}
R.~S. Anderson and P.~K. Haff, Science {\bf 241},  820  (1988).

\bibitem{ActaMechanica91}
{\em Aeolian Grain Transport 1 {\&} 2}, {\em Acta Mechanica Suppl.}, edited by
  O.~E. Barndorff-Nielsen and B.~B. Willetts (Springer, Wien, New York, 1991).

\bibitem{logs}
  As long as $L$ and $z_0$ do not change by
  orders of magnitude, this weak logarithmic dependence is
  negligible.

\bibitem{finite_width}
  Analytic evaluation for a dune with a Gaussian transverse
  profile of width $\sigma$ shows that the main effect of a finite
  transverse width is to renormalize $\beta\to\beta(\sigma)
  \small{\lesssim} \beta$ and to add a (trivial) term $\gamma(\sigma)
  h(x)$ with $\gamma( \infty)=0$ and $\gamma( L/\sqrt2) \approx -0.2$
  within the brackets of Eq.(\ref{eq:wind_model}).

\bibitem{lorentz}
For the contour $f_L$, the relative shift $\delta x/\! L$ of the maximum of
  $\tau(x)$ with respect to the maximum of $f(x)$ is $2\,(1 +
  \beta^2)^{1/2}\,\sin [\arctan (\beta)/3] -\beta \approx -\beta/3$ 
 ($\beta\ll1$).

\bibitem{Wippermann86}
F.~K. Wippermann and G. Gross, Boundary Layer Meteorology {\bf 36},  319
  (1986).

\bibitem{van_dijk-arens-boxel:99}
P.~M. van Dijk, S.~M. Arens, and J.~H. van Boxel, Earth Surf. Process.
  Landforms {\bf 24},  319  (1999).

\bibitem{zeman-jensen:88}
O.~Zeman and N.~O.~Jensen, Ris{\o} National Laboratory {\bf M-2738},    (1988).

\bibitem{Stam97}
J.~M.~T. Stam, Sedimentology {\bf 44},  127  (1997).

\bibitem{bubble} 
We have carefully tested this heuristic procedure
against $2d$ and $3d$ numerical solutions from the finite volume solver FLUENT5
(Fluent Inc.\ 1999) and verified that the predictions for the windward
side are reasonably stable against the freedom thus introduced into
the model.

\bibitem{sauermann-etal:2000}
G. Sauermann, P. Rognon, A. Poliakov, and H.~J. Herrmann, Geomorphology {\bf
  36},  47  (2000).

\bibitem{Wiggs96}
G.~F.~S. Wiggs, I. Livingstone, and A. Warren, Geomorphology {\bf 17},  29
  (1996).

\bibitem{Hesp98}
P.~A. Hesp and K. Hastings, {Geomorphology} {\bf 22},  193  (1998).

\bibitem{sykes:80}
R.~I. Sykes, J.~Fluid Mech. {\bf 101},  647  (1980).

\bibitem{zeman-jensen:87}
O.~Zeman and N.~O.~Jensen, Q. J. R. Meteorol. Soc. {\bf 113},  55  (1987).

\bibitem{boxel-arens-van_dijk:99}
J.~H. van Boxel, S.~M. Arens, and P.~M. van Dijk, Earth Surf. Process.
  Landforms {\bf 24},  255  (1999).

\end{thebibliography}

\end{multicols}

\end{document}